\documentclass{article}
\usepackage[T1]{fontenc}
\usepackage[utf8]{inputenc}
\usepackage[lbd]{ismir}
\usepackage{amsmath,cite,url}
\usepackage{graphicx}
\usepackage{color}
\usepackage{booktabs} % \toprule, \midrule, \bottomrule
\usepackage{makecell} % \makecell
\usepackage{multirow}

\title{CloudEar: Integrating Perceptual, Diagnostic, and Cross-Modal Evidence for Music Evaluation}

\twoauthors
  {Qiqi He}
  {NetEase Cloud Music\\
   {\tt heqiqi02@corp.netease.com}}
  {Anqi Huang}
  {NetEase Cloud Music\\
   {\tt huanganqi01@corp.netease.com}}

\def\authorname{Q. He and A. Huang}

\begin{document}

\maketitle

\begin{abstract}
Evaluating generated music requires modeling perceptual quality, technical
audio quality, and prompt alignment. Existing methods often overlook
signal-level defects, produce compressed quality scores, and miss
fine-grained musical attributes in text prompts. We propose \textit{CloudEar},
a pairwise framework for evaluating musicality and prompt alignment.
It comprises three components: \textit{i)} a Perceptual Expert for song-quality assessment;
\textit{ii)} a Diagnostic Expert for learned and signal-level quality
analysis; and \textit{iii)} a Cross-modal Expert for prompt--song alignment based on musical attributes. Experiments show that \textit{CloudEar}
outperforms the compared baselines in pairwise preference accuracy
and score correlation.
\end{abstract}

\section{Introduction}

Evaluating generated music requires assessing both audio quality
and musical aesthetics. Signal-level defects such as clipping and abnormal
loudness can degrade listening quality, while musical structure, expression,
and performance require higher-level perceptual judgment
~\cite{sloboda1985musical}. Existing methods emphasize high-level musical semantics but often overlook
technical quality and subtle perceptual differences. Regression models also
exhibit central-tendency bias~\cite{regression_bias}, resulting in compressed
score differences.

In parallel, prompt alignment poses a separate challenge. General-purpose
audio--text models capture broad semantic correspondence but often overlook
fine-grained musical attributes such as instrumentation requirements.
Consequently, high audio--text similarity does not always indicate whether a
generated song follows the specific attributes requested by the prompt.

To address these limitations, we propose \textit{CloudEar}, a pairwise framework
for evaluating musicality and prompt alignment. Its Perceptual Expert combines
ordinal regression, continuous regression, and pairwise ranking to model
human-perceived song quality to mitigate prediction concentration. A
Diagnostic Expert complements this representation with learned quality
estimates and signal-level descriptors. For alignment, a Cross-modal Expert
compares genre and instrumentation attributes extracted independently from
the audio and prompt. Experiments on rating and pairwise-preference
benchmarks show that \textit{CloudEar} consistently improves both musicality and
alignment evaluation over existing methods.
\section{Proposed Framework}
\label{sec:framework}
\begin{figure}[t]
    \centering
    \includegraphics[width=\linewidth]{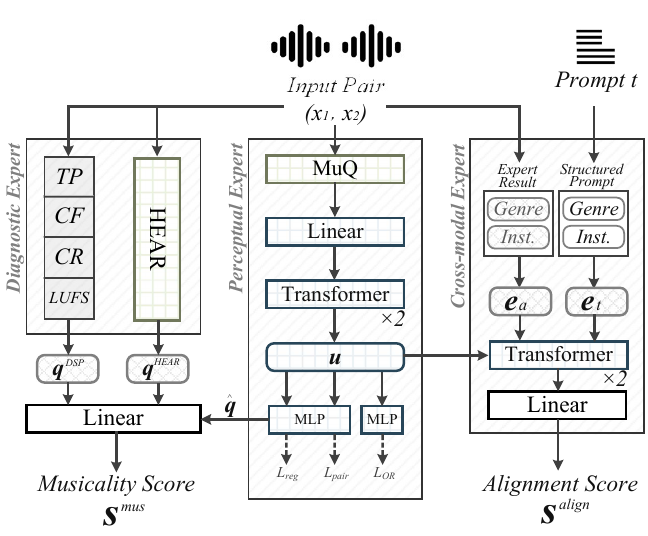}
    \caption{Overview of the proposed \textit{CloudEar} framework.}
    \label{fig:pipeline}
\end{figure}

As illustrated in Fig.~\ref{fig:pipeline}, given a prompt $t$ and two
generated songs $(x_1,x_2)$, \textit{CloudEar} predicts their pairwise
preferences in musicality and prompt alignment.

\subsection{Perceptual Expert}
\label{sec:perceptual_experts}

Following SongBench~\cite{songbench}, we randomly sample a 30-second clip
from each song $x_i$ and extract MuQ features~\cite{muq}, which are
processed by a Transformer adapter to obtain the perceptual
representation $u_i$.

To alleviate the tendency of regression models to concentrate predictions
around the mean~\cite{regression_bias}, we jointly optimize ordinal
regression, continuous regression, and pairwise ranking objectives.
The ordinal loss $L_{\mathrm{OR}}$~\cite{or_loss} supervises ordered
quality levels, while an MLP regression head predicts the quality score
$\hat{q}_i$ using $L_{\mathrm{reg}}=(\hat{q}_i-y_i)^2$,
where $y_i$ is the ground-truth score.
For a training pair with $y_1>y_2$, the pairwise loss encourages
a higher score for the preferred song:
\begin{equation}
L_{\mathrm{pair}}
=
-\log \sigma(\hat{q}_1-\hat{q}_2),
\end{equation}
where $\sigma$ denotes the sigmoid function.
The overall training objective is
\begin{equation}
L_{\mathrm{PE}}
=
0.4L_{\mathrm{OR}}
+
0.4L_{\mathrm{reg}}
+
0.2L_{\mathrm{pair}}.
\end{equation}

\subsection{Diagnostic Expert}
\label{sec:audio_diagnostics}

The Diagnostic Expert complements high-level perceptual modeling with
signal-level features and learned quality estimates. We extract
true-peak level ($\mathrm{TP}$), clipping ratio ($\mathrm{CR}$),
crest factor ($\mathrm{CF}$), and integrated loudness
($\mathrm{LUFS}$) to form the DSP feature vector
$\mathbf{q}_i^{\mathrm{DSP}}$, with loudness and true-peak measurements
following standard practices~\cite{itu_bs1770,ebu_r128}.
We also use HEAR~\cite{hear}, which combines MuQ and MusicFM
representations, to obtain the quality score $q_i^{\mathrm{HEAR}}$.
The DSP feature vector, HEAR score, and perceptual score $\hat{q}_i$
are concatenated and fed into a linear layer to produce the
musicality score $s_i^{\mathrm{mus}}$.

The musicality fusion layer is trained using binary cross-entropy
(BCE) between the pairwise preference label and the predicted
probability $\sigma(s_1^{\mathrm{mus}}-s_2^{\mathrm{mus}})$.
At inference time, \textit{CloudEar} selects the song with the higher
musicality score.

\subsection{Cross-modal Expert}
\label{sec:cross_modal_alignment}

The Cross-modal Expert assesses genre and instrument alignment between the prompt and song. On the audio side, Genre Discogs400~\cite{alonso2023efficient} predicts
400 music styles, and MTG-Jamendo Instrument~\cite{bogdanov2019mtg}
predicts instrument tags. Their outputs are concatenated into $e_{a,i}$. On the prompt side, we use Qwen3-30B~\cite{qwen3} to extract structured genre and instrument information from $t$. The extracted attributes are
mapped to label vectors using the same vocabularies as the audio classifiers. These vectors are concatenated to form
the prompt representation $e_t$.

A two-layer Transformer jointly models $u_i$, $e_{a,i}$, and $e_t$,
followed by a linear layer that produces the alignment score
$s_i^{\mathrm{align}}$. The alignment branch is trained with BCE
on pairwise preference probabilities obtained by applying a sigmoid
to the score differences.
For each pair conditioned on the same prompt, \textit{CloudEar} selects
the song with the higher alignment score.

\section{Experiments}

\subsection{Datasets \& Metrics}
We train the Perceptual Expert on \texttt{AIME}~\cite{aime},
\texttt{MusicArena}~\cite{musicarena}, and approximately 7,000 hours of
in-house human-rated music, mainly English songs from five generation models.
For the in-house dataset, we retain only songs rated by more than 200 listeners.

For pairwise training, we additionally use \texttt{MusicEval}~\cite{musiceval} and the training split
of \texttt{CMI-Pref}~\cite{cmi_rm}. Their musicality annotations train the
Perceptual Expert and musicality fusion layer, while their alignment
annotations train the Cross-modal Expert. Musicality pairs use within-dataset rating gaps above 0.2 after $[0,1]$
normalization while alignment pairs share a prompt.

Following prior work~\cite{cmi_rm}, evaluation uses
\texttt{PAM-Music}~\cite{pam}, \texttt{CMI-Test}, and the
\texttt{MusicArena} test set (MA.): 1,340 song pairs generated
in Jul.--Dec. 2025, with no training-data overlap. We report
SRCC and LRCC for scalar ratings and P-ACC for A/B preferences.
\subsection{Experimental Results}

\textbf{Ablation Study.}
The full \textit{CloudEar} outperforms all reduced configurations on musicality, confirming their complementary contributions. Both HEAR and DSP improve over \textit{PE only}, demonstrating complementary information beyond the perceptual representation. The alignment gain over \textit{PE only}, trained directly on alignment labels, supports the benefit of the Cross-modal Expert. The Perceptual Expert alone already outperforms SongBench on \texttt{CMI-Test} musicality and remains competitive on \texttt{MusicArena-Test}.

\textbf{Comparison with Prior Work.}
For baseline evaluation, SongBench and HEAR use musicality scores for both tasks. \textit{CloudEar} outperforms all compared baselines on the reported
benchmarks. Its higher musicality and alignment P-ACC indicate better agreement with human judgments of quality and prompt alignment, respectively. Higher SRCC and LRCC on \texttt{PAM-Music} further demonstrate stronger rank and
linear correlations with human quality ratings. Notably, SongBench and HEAR achieve above-chance alignment accuracy without prompts, suggesting a correlation between musicality and alignment preferences.

\begin{table}[t]
\centering
\caption{Comparison and ablation results. Mus./Ali.: musicality/alignment.
MuQ uses MuQ-Eval~\cite{muqeval} for Mus. and MuQ-MuLan\cite{muq} audio--text similarity for Ali.}
\label{tab:main_results}
\small
\setlength{\tabcolsep}{3.5pt}
\renewcommand{\arraystretch}{0.95}
\begin{tabular}{l|cc|c|cc}
\toprule
\multirow{2}{*}{Method}
& \multicolumn{2}{c|}{\texttt{CMI-Test}}
& MA.
& \multicolumn{2}{c}{\texttt{PAM-Music}} \\
& \makecell{Mus.\\P-ACC}
& \makecell{Ali.\\P-ACC}
& P-ACC
& SRCC
& LRCC \\
\midrule
CMI-RM~\cite{cmi_rm}
& .7833 & .7389 & .7194 & .6626 & .6379 \\
HEAR~\cite{hear}
& .7882 & .6970 & .5769 & .6180 & .6230 \\
SongBench~\cite{songbench}
& .7340 & .6823 & .6746 & .5437 & .5123 \\
MuQ~\cite{muq, muqeval}
& .5936 & .6429 & .6478 & .3168 & .3608 \\
\midrule
\textit{CloudEar}
& \textbf{.8179} & \textbf{.7635} & \textbf{.7657}
& \textbf{.7257} & \textbf{.7229} \\
\textit{PE + HEAR}
& .7805 & -- & .7325 & .6872 & .7008 \\
\textit{PE + DSP}
& .7620 & -- & .6845 & .6351 & .6626 \\
\textit{PE only}
& .7592 & .6743 & .6736 & .6056 & .6515 \\
\bottomrule
\end{tabular}
\end{table}
\vspace{-.2cm}
\section{Conclusion}
We presented \textit{CloudEar}, a pairwise framework for evaluating the
musicality and prompt alignment of generated songs. \textit{CloudEar} integrates perceptual, diagnostic, and cross-modal experts to
model song quality and fine-grained prompt alignment. Experiments on three
human-annotated benchmarks demonstrate consistent improvements over existing
methods. Future work will explain which musical factors lead to each preference.
\section{AI Usage Statement}
Qwen3-30B is used to extract structured genre and instrument
attributes from input prompts, as described in
Section~\ref{sec:cross_modal_alignment}.
The authors take full responsibility for the content of this paper.
\bibliography{references}
\end{document}